\documentclass[aps,twocolumn,superscriptaddress,prl,bibnotes]{revtex4}

\usepackage{graphicx}

\begin{document}


\title[]{Anisotropic modification of the effective hole $g$-factor by electrostatic confinement}

\author{S.~P. Koduvayur}
\email[]{sunanda@purdue.edu}
\author{L.~P. Rokhinson}

\affiliation{Department of Physics, Purdue University, West Lafayette, IN 47907
USA}

\author{D.~C. Tsui}
\affiliation{Department of Electrical Engineering, Princeton University,
Princeton, NJ 08544 USA}

\author{L.~N. Pfeiffer}
\author{K.~W. West}
\affiliation{Bell Laboratories, Lucent Technologies, Murray Hill, New Jersey
07974 USA}


\begin{abstract}
We investigate effects of lateral confinement on spin splitting of energy
levels in 2D hole gases grown on [311] GaAs. We found that lateral confinement
enhances anisotropy of spin splitting relative to the 2D gas for both confining
directions. Unexpectedly, the effective $g$-factor does not depend on the 1D
energy level number $N$ for $B\|[0\overline{1}1]$ while it has strong
$N$-dependence for $B\|[\overline{2}33]$. Apart from quantitative difference in
the spin splitting of energy levels for the two orthogonal confinement
directions we also report qualitative differences in the appearance of
spin-split plateaus, with non-quantized plateaus observed only for the
confinement in $[0\overline{1}1]$ direction. In our samples we can clearly
associate the difference with anisotropy of spin-orbit interactions.
\end{abstract}


\maketitle

Devices that use spin as the main carrier of information promise higher speeds
and lesser energy demands and have been the bases for the new fields of
spintronics and quantum information \cite{wolf01,zutic04}. An important aspect
in the realization of these devices is efficient manipulation and control of
spins. GaAs hole systems provide a potential advantage in electrostatic
manipulation of spins due to stronger spin-orbit (SO) interaction, compared to
electronic systems. With predictions of increasing spin-relaxation times in
p-type based low-dimensional systems \cite{loss98} to orders comparable to
those of electrons, there is a need to better understand the physics of SO
interactions.

In two-dimensional GaAs hole gases (2DHG) grown in [001] crystallographic
direction, SO locks spins in the growth direction resulting in a vanishing spin
response to the in-plane magnetic field (vanishing effective Land\'{e}
$g$-factor $g^*$)\cite{lin91,rahimi03}. For high-index growth directions, such
as [311], in-plane $g^{*}$ is not zero and becomes highly
anisotropic\cite{winkpap}. Additional lateral confinement increases $g^{*}$
anisotropy\cite{danneau06} and the value depends on the population of 1D
subbands\cite{daneshvar97}. Strong suppression of $g^{*}$ for the in-plane
magnetic field perpendicular to the channel direction has been attributed to
the confinement-induced re-orientation of spins perpendicular to the 1D
channel\cite{danneau06}. In this Letter we demonstrate that the anisotropy of
spin splitting is primarily due to the crystalline anisotropy of SO
interactions and not the lateral confinement. We investigate quantum point
contacts with confinement in both $[0\overline{1}1]$ and $[\overline{2}33]$
directions and find that anisotropy of spin splitting depends on the field
direction rather than on the direction of the lateral confinement. There is a
strong dependence of $g^{*}$ on the number of filled 1D subbands $N$ for one
field direction ($B\|[\overline{2}33]$), while $g^{*}$ is almost
$N$-independent for the orthogonal field direction ($B\|[0\overline{1}1]$). We
also report qualitative differences in the appearances of the conductance
plateaus for the two orthogonal confinement directions. For the channels
confined in $[\overline{2}33]$ direction the conductance of spin-split plateaus
is $(N+1/2)e^2/h$, in accordance with Landauer formula. For the orthogonal
direction non-quantized plateaus appear that have some resemblance to the
so-called ``0.7 structure''\cite{thomas96} and its various
``analogs''\cite{graham03} and their conductance values change with magnetic
field. The major difference between the two orientations of 1D channels in our
experiments is the strength of SO, which may provide some clues to the origin
of these yet-to-be-understood anomalies.

We use AFM local anodic oxidation\cite{snowheld} to fabricate the QPCs, which
results in sharper potential compared to top gating technique and also
eliminates leakage problems associated with low Schottky barriers in p-GaAs.
The use of this technique requires specially designed heterostructures with
very shallow 2DHG,details of which are given in \cite{rokhinson02}. An AFM
image of a QPC device is shown in the inset in Fig.~\ref{confinf}. White lines
are oxide which separates 2DHG into source (S), drain (D) and gate (G) regions,
the 2DHG is depleted underneath the oxide. The side gates are used to
electrostatically control the width of the 1D channel. AFM lithography aids in
precise control of QPC dimensions with corresponding pinch-off voltage control
within a few mV, allowing comparison of orthogonal QPCs with similar confining
potential. At $T=4$ K, QPCs show regular smooth FET characteristics as a
function of gate voltage. For orthogonal QPCs with similar pinch-off voltages,
resistances differ by a factor of two, reflecting the underlying anisotropy of
the 2DHG. Conductivity of 2DHG on [311] GaAs is anisotropic due to a
combination of effective mass anisotropy and difference in surface morphology,
with $[\overline{2}33]$ being high-mobility and $[0\overline{1}1]$ low-mobility
directions\cite{heremans94}.

\begin{figure}[t]
\def\ffile{confinf}
\hspace{-0.5in}
\includegraphics[scale=0.11]{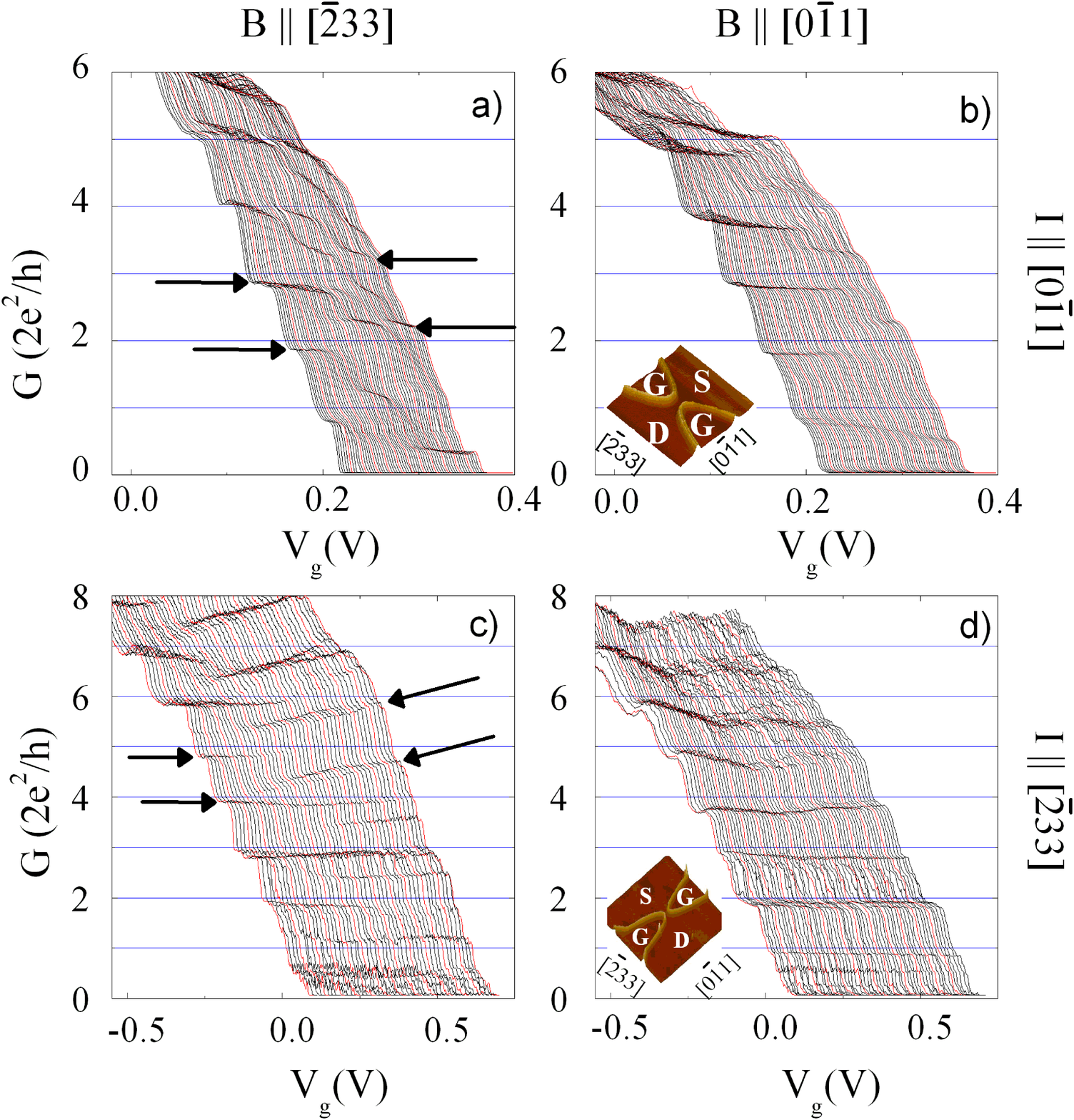}
\caption{Conductance of QPCs as a function of gate voltage. The curves are
offset proportional to $B$ with 0.25T interval. Leftmost curve corresponds to
$B=0$. (a,b) are for the channel along $[0\overline{1}1]$ and (c,d) for the
channel along $[\overline{2}33]$. The arrows highlight a few plateaus discussed
in the text, the slope of the arrows highlighting the slope of the
corresponding plateau. Insets: $2\mu m \times 2\mu m$ AFM micrographs of
devices.}
\label{\ffile}
\end{figure}

Typical conductance of QPCs at low temperatures is shown in Figs.~\ref{confinf}
and~\ref{highmuIf}. Leftmost curves are measured for $B=0$. Four-terminal
resistance is corrected for the gate-independent series resistance of the
adjacent 2D gas, $R_{0}=300-600 \Omega$ in different samples. $R_{0}$ was also
corrected for its $B$-dependence which was measured separately for both
crystallographic directions (a 20\% increase at 12T). For the sample studied in
Figs.~\ref{confinf} (a,b) the 1D channel is confined in [$\overline{2}33$]
direction ($I\| [0\overline{1}1]$), while in Figs.~\ref{confinf} (c,d) and
~\ref{highmuIf} (a,b) it is confined in [$0\overline{1}1$] direction
($I\|[\overline{2}33]$). At low temperatures conductance is
quantized\cite{wharam88,vanwees88,zailer94} in units of $G=Ng_{0}$, where
$g_{0}=2e^2/h$ and $N$ is the number of 1D channels below the Fermi energy,
which reflects the exact cancellation of carriers velocity and the density of
states in 1D conductors. The factor 2 reflects spin degeneracy of energy levels
at $B=0$. Plateaus appear when electrochemical potentials of source and drain
lie in the gap between neighboring 1D subbands $E^N$ and $E^{N+1}$. In various
samples we resolve up to 10 plateaus at temperatures $T<100$ mK.

Effect of in-plane magnetic field on conductance is shown in
Figs.~\ref{confinf} and~\ref{highmuIf} for the two orthogonal field direction.
The curves are offset proportional to the magnetic field with 0.25T increments.
The samples were rotated either {\it in situ} (Fig.~\ref{highmuIf}) or after
thermo-cycling to room temperature (Fig.~\ref{confinf}). Mesoscopic changes
during thermo-cycling are reflected in a small difference between the $B=0$
curves, yet they do not change level broadening and onset of spin splitting
significantly.

\begin{figure}[t]
\def\ffile{highmuIf}
\includegraphics[scale=0.13]{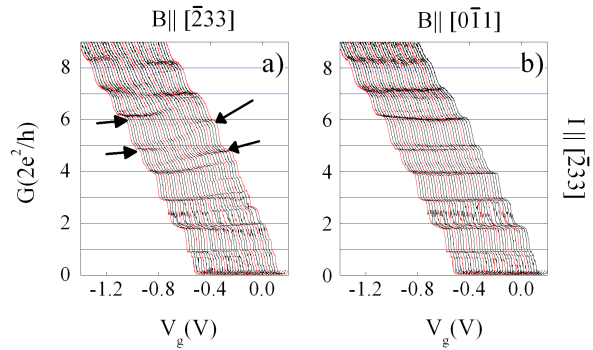}
\caption{Conductance of another QPC with the channel oriented along
$[\overline{2}33]$. The curves are offset proportional to $B$ with 0.25 $T$
interval. Leftmost curve corresponds to $B=0$. The arrows highlight plateaus
discussed in the text.}
\label{\ffile}
\end{figure}

There are both quantitative and qualitative differences in the field response
of orthogonally oriented 1D channels. We begin the analysis with a quantitative
comparison of spin splitting of energy levels for different orientations of
magnetic field and channel directions. In general the energy spectrum for holes
contains linear, cubic and higher-order terms in $B$\cite{winkler03}. At low
fields the linear term dominates and we approximate spin splitting by the
Zeeman term with an effective $g$-factor, $E_Z=2g^*_{[ijk],N} \mu_B B$, where
$\mu_B$ is the Bohr magneton and $g^*_{[ijk],N}$ depends on field orientation
$B\|[ijk]$, energy level number $N$ and confinement direction. Half-integer
plateaus appear at the critical fields $B^{N-1/2}$, when spin splitting of the
$N$-th level becomes equal to the disorder broadening of the level, as shown
schematically in Fig.~\ref{whitegrayf}(d). While level broadening is different
for different energy levels we expect it to be independent of the direction of
the magnetic field and hence the ratio of $g^*$'s for the two orthogonal
directions can be obtained from the appearance of half-integer plateaus,
$B^{N-1/2}_{[1\overline{1}0]}/B^{N-1/2}_{[\overline{2}33]}=g^{*}_{[\overline{2}33],N}/g^{*}_{[1\overline{1}0],N}$.
The integer plateaus disappear at the fields $B^{N}$ when two neighboring
levels with opposite spin intersect, and the average $\langle
g^*_{[ijk],N}\rangle=(g^{*}_{[ijk],N}+g^{*}_{[ijk],N+1})/2$ can be found from
$\Delta E_{N}=\Delta E_{z}=\langle g^*_{[ijk],N}\rangle\mu _{B}B^{N}_{[ijk]}$,
where $\Delta E_{N}$ is the zero-field energy spacing of 1D subbands excluding
level broadening.

\begin{figure}[t]
\def\ffile{whitegrayf}
\includegraphics[scale=0.11]{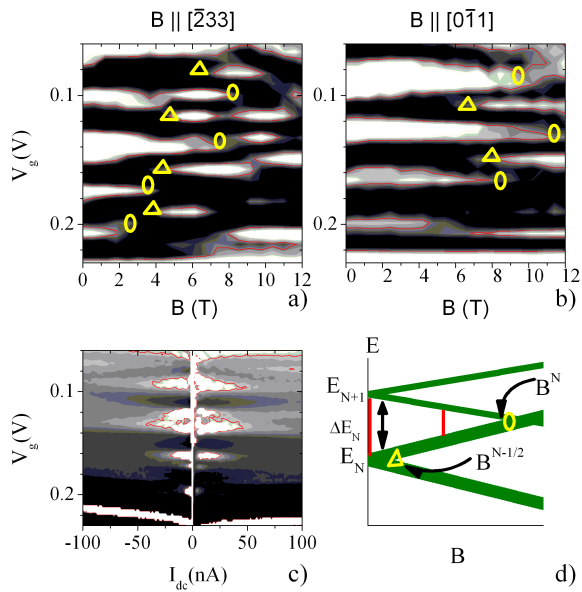}
\caption{(a,b) Derivative of curves in Fig~\ref{confinf}(a,b), white regions
correspond to the conductance plateaus. (c) Differential transresistance
plotted in a logarithmic scale (from 0.01 $k\Omega$ (white) to 0.2 $k\Omega$
(black)) for the same sample at $B=0$. (d) Schematic of Zeeman splitting of
energy levels.}
\label{\ffile}
\end{figure}

\begin{figure}[t]
\def\ffile{gNff}
\includegraphics[scale=0.4]{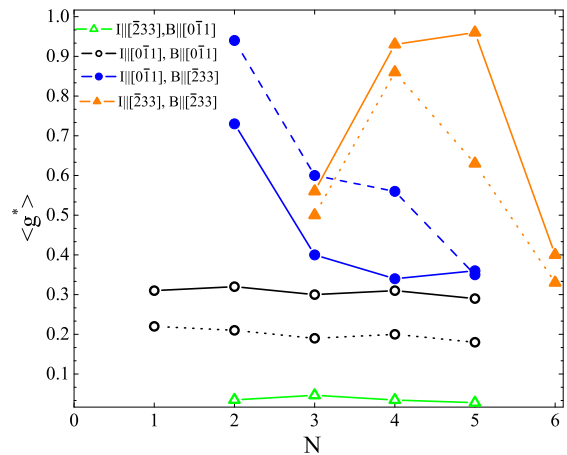}
\caption{(a) Average $g_{N}^{*}$ between adjacent levels $N$ and $N+1$ is
plotted for different orientations of channel and magnetic field. Open and
filled symbols are for magnetic field parallel to $[0\overline{1}1]$ and
$[\overline{2}33]$, respectively. Circles and triangles are for channels along
$[0\overline{1}1]$ and $[\overline{2}33]$, respectively. The blue dashed curve
is the actual $g^*_{N}$ for $I\|[0\overline{1}1],B\|[\overline{2}33]$. The
orange and black dotted curves are corrected for the diamagnetic shift.}
\label{\ffile}
\end{figure}

Splitting and crossing of energy levels are best visualized in transconductance
plots. In Fig.~\ref{whitegrayf}(a,b) a grayscale of $dG/dV_g$ for the data in
Fig.~\ref{confinf}(a,b) is plotted. The white regions correspond to the
plateaus, the dark regions correspond to the energy level being aligned with
the Fermi energy in the leads and reflect level broadening, which is roughly
half of the level spacing in our samples. At low fields the width of the
plateaus decreases almost linearly with field, hence justifying the use of
linear approximation but at high fields there is a clear deviation from linear
dependence. The critical fields where levels cross ($B^{N}$) and split
($B^{N-1/2}$) are indicated by triangles and circles.

Level spacing is determined from non-linear transport spectroscopy. A
logarithmic scale plot of transconductance for the same sample is shown in
Fig.~\ref{whitegrayf}(c) with white regions representing the plateaus. By
determining the maximum current $I_{max}$ for the $N^{th}$ plateau at which the
transconductance is still zero we obtain the 1D subband spacings between levels
$N$ and $N+1$ (excluding level broadening)as $\Delta E_{N}=eRI_{max}$, where
$R=h/2Ne^{2}$ is the resistance on the plateaus.

\begin{table*}
\begin{tabular}
{||c||c||c|c||c|c|c|c||c|c||} \hline $N$&$R$(k$\Omega)$&$I_{max}$(nA)&$\Delta
E_N$($\mu$eV)& $B^{N}_{[\overline{2}33]} $(T)&$\angle
g^*_{[\overline{2}33],N}\rangle$& $B^{N}_{[0\overline{1}1]} $(T)&$\langle
g^*_{[0\overline{1}1],N}\rangle$&
$\frac{B^{N-1/2}_{[0\overline{1}1]}}{B^{N-1/2}_{[\overline{2}33]}}=\frac{g^*_{[\overline{2}33]}}{g^*_{[0\overline{1}1]}}$&
$g^*_{[\overline{2}33]}$\\
\hline\hline
1&12.9&6&80&&&4.5&0.31&&\\
2&6.45&23&150&3.6&0.73&8&0.32&3&0.94\\
3&4.3&40&170&7.5&0.4&10&0.30&2&0.6\\
4&3.225&50&160&8&0.34&9&0.31&1.8&0.56\\
5&2.58&60&150&7.3&0.36&9&0.29&1.2&0.35\\
\hline
\end{tabular}
\caption{Summary of experimental values used to extract $g^*$ for different
energy levels for channel along $[0\overline{1}1]$.}
\label{table1}
\end{table*}
\begin{table*}
\begin{tabular}{||c||c||c|c||c|c||c||c|c||}
\hline $N$&$R$(k$\Omega)$&$I_{max}^{B=0}$(nA)&$\Delta
E_N^{B=0}$($\mu$eV)&$I_{max,[0\overline{1}1]}^{8T}$(nA)&$\Delta
E_{N,[0\overline{1}1]}^{8T}$($\mu$eV)&$\langle
g^*_{[0\overline{1}1],N}\rangle$&$B^{N}_{[\overline{2}33]} $(T)&$\langle
g^*_{[\overline{2}33],N}\rangle$
\\
\hline\hline
2&6.45&25&161.25&22.5&145.13&0.035&&\\
3&4.3&27.5&118.25&22.5&96.75&0.046&3&0.56\\
4&3.225&50&161.25&45&145.13&0.0347&3&0.93\\
5&2.58&42.5&109.65&37.5&96.75&0.028&6&0.96\\
6&2.16&35&75.6&&&&3.25&0.4\\
 \hline
\end{tabular}
\caption{Summary of experimental values used to extract $\langle g^*\rangle$
for different energy levels for channel along $[\overline{2}33]$}.
\label{table2}
\end{table*}


The experimental data for the channel along $[0\overline{1}1]$ is summarized in
Table~\ref{table1}. We obtain the energy level spacing $\Delta E_{N}$ for the
first five energy levels using the above explained method. From the critical
fields $B^{N}$ we obtain the average $\langle g^{*}_{[ijk],N}\rangle$ for the
neighboring energy levels. The ratio of the $g^{*}$s is 3 for $N=1$ and
approaches the 2D value of 1.2 for large $N$. The values $\langle
g^*_{[0\overline{1}1]}\rangle$ do not depend on $N$ and we use
$g^*_{[0\overline{1}1]}=0.3$ to obtain the values for
$g^{*}_{[\overline{2}33]}$ from the ratios
$g^{*}_{[\overline{2}33],N}/g^{*}_{[1\overline{1}0],N}$.  In Table~\ref{table2}
we present similar data for QPCs with the channel along $[\overline{2}33]$
direction. For these samples no half-split plateaus are observed for $B \|
[0\overline{1}1]$ and $B^{N-1/2}$ is unattainable. We still can extract the
average $\langle g^{*}\rangle$ values by measuring the change in the energy
level spacing $\Delta E_{N}(0)-\Delta E_{N}(B)=\langle g^{*}\rangle \mu _{B}
B$, as shown by bars in the schematic in Fig.~\ref{whitegrayf}(d). For
$B\|[\overline{2}33]$ the introduction of $g^{*}$ has questionable meaning due
to anomalous behavior of half-integer plateaus and ill-defined $B^{N-1/2}$. We
estimate $g^*$ from measured $B^N$.

Fig.~\ref{gNff} summarizes our results for the $g^{*}$ for different
confinement directions. For $B\|[\overline{2}33]$ spin splitting of energy
levels strongly depends on the level number $N$ for both confinement
directions. For the field $B\|[0\overline{1}1]$, $g^*$  is smaller and is
almost independent of $N$. We see this trend for all the four samples we
measured. We conclude that $g$-factor anisotropy is primarily determined by the
crystalline anisotropy of spin-orbit interactions. Lateral confinement enhances
the anisotropy.

So far we ignored diamagnetic shift of energy levels. The ratios
$g^{*}_{[\overline{2}33],N}/g^{*}_{[1\overline{1}0],N}$ are not affected by
this shift because they characterize energy difference between spin states of
the same orbital level. Likewise, the extracted $\langle g^{*}\rangle$ will not
be affected by field confinement in the growth direction because the first 8-10
1D levels belong to the same lowest 2D subband. The only value to be affected
by diamagnetic shift will be $\langle g^{*}\rangle$ for $B\|I$. To estimate the
correction we approximate both vertical and lateral confinement by parabolic
potentials $\hbar\omega_{z}=2.4$ meV , $\hbar\omega_{y}=0.3$ meV. The corrected
$\langle g^{*}_{c}\rangle=\langle
g^{*}\rangle(1+\frac{\omega_{1}(B^{N})-\omega_{1}(0)}{\omega_{1}(0)})$, where
$\hbar\omega_{1}=\frac{\hbar}{2}\sqrt{(\omega_{c}^{2}+\omega_{y}^{2}+
\omega_{z}^{2})-\sqrt{(\omega_{c}^{2}+\omega_{y}^{2}+\omega_{z}^{2})^{2}
-\omega_{y}^{2}\omega_{z}^{2}}}$ is the field dependent energy spacing for
spinless particles\cite{Salis99}, $\omega_{c}=eB/m_c$ is the cyclotron
frequency, and $m_{c}=\sqrt{m_{h}m_{l}}=0.28m_{e}$ is the cyclotron mass. For
$I\| [\overline{2}33]$ the critical fields $B^N_{[\overline{2}33]}\sim 3T$ are
small and correction to $\langle g^{*}\rangle$ due to diamagnetic shift is
$<5\%$. For the channel along $[0\overline{1}1]$ $B^N_{[0\overline{1}1]}\sim
8-10$T and correction is $\sim 30\%$ which is not negligible. We plot the
corrected values in Fig.~\ref{gNff}.

Now we highlight a few qualitative differences in the appearance of
"half-integer" plateaus for the channels along $[0\overline{1}1]$ and
$[\overline{2}33]$ directions. Conductance of spin-split plateaus for channels
along $[0\overline{1}1]$ are quantized at $G=(N+1/2)g_0$, in full agreement
with the theory. In point contacts with confinement in the orthogonal direction
conductance of spin-split plateaus is not quantized and is field dependent. At
low fields ($B<4$T) their evolution resembles ``0.7 structure'' and various
anomalous plateaus reported in electron samples. At higher fields the
conductance of these plateaus increases with magnetic field, at the same time
the integer plateaus remain quantized at $Ng_{0}$. We emphasize the motion of
spin-split plateaus with the slope of arrows in Figs.~\ref{confinf}
and~\ref{highmuIf}. For example, in Fig.~\ref{confinf}(c) a plateau at
4.3$g_{0}$ appears at $B\sim 3$T and its value gradually increases to $\sim 4.8
g_{0}$ by $12$T. The next non-integer plateau appears at $B\sim 3$T and
increases to $\sim6g_{0}$ by $B=12$T, while the neighboring integer plateaus
remain quantized at $G=4g_{0},5g_{0}$ and $6g_{0}$. This feature has been
observed consistently in all the samples we measured, as is evident from
Fig.~\ref{highmuIf}, where similar data is presented for a different sample: a
plateau at 5.2$g_{0}$ appears at $B\sim3.3$T and increases to $\sim6g_{0}$ by
$8$T. The orthogonal 1D channels are fabricated from the same 2D hole gas and
have similar confinement potentials. The only difference is due to the
anisotropy of spin-orbit interactions. Thus, we conclude that spin-orbit
interactions are responsible for the anomalous behavior.

To summarize the results, we investigate effects of lateral confinement on spin
splitting of energy levels in 2D hole gases in [311]GaAs. We found that lateral
confinement enhances anisotropy of spin splitting relative to the 2D gas for
both confining directions. Unexpectedly, the effective $g$-factor does not
depend on the energy level number $N$ for $B\|[0\overline{1}1]$ while it has
strong $N$-dependence for the orthogonal orientation, $B\|[\overline{2}33]$. We
also observe qualitative differences in the appearance of spin-split plateaus
for the two orthogonal directions of lateral confinement, which we can
attribute to the difference in spin-orbit interaction.

This work was supported by NSF grant ECS-0348289.

\bibliographystyle{revtex-etal}
\bibliography{rohimod,g_factor}

\begin{thebibliography}{10}

\bibitem{wolf01}
S.~A. Wolf {\it et~al.}, Science {\bf 294},  1488  (2001).

\bibitem{zutic04}
I. Zutic, J. Fabian, and S. Das~Sarma, Rev. Mod. Phys. (USA) {\bf 76},  323
  (2004).

\bibitem{loss98}
D. Loss and D.~P. DiVincenzo, \pra {\bf 57},  120  (1998).

\bibitem{lin91}
S.~Y. Lin {\it et~al.}, Phys. Rev. B {\bf 43},  12110  (1991).

\bibitem{rahimi03}
M. Rahimi {\it et~al.}, Phys. Rev. B {\bf 67},  081302(R)  (2003).

\bibitem{winkpap}
S.~J. Papadakis, E.~P. De~Poortere, M. Shayegan and R. Winkler, Phys. Rev.
  Lett. \textbf{84}, 5592 (2000); R. Winkler, S. Papadakis, E.~P. De~Poortere,
  and M. Shayegan, {\it ibid.}, \textbf{85}, 4574 (2000).

\bibitem{danneau06}
R. Danneau {\it et~al.}, Phys. Rev. Lett. {\bf 97},  026403   (2006).

\bibitem{daneshvar97}
A.~J. Daneshvar {\it et~al.}, Phys. Rev. B {\bf 55},  R13409  (1997).

\bibitem{thomas96}
K.~J. Thomas {\it et~al.}, Phys. Rev. Lett. {\bf 77},  135  (1996).

\bibitem{graham03}
A.~C. Graham {\it et~al.}, Phys. Rev. Lett. {\bf 91},  136404  (2003).

\bibitem{snowheld}
E. S. Snow and P. M. Campbell, Science \textbf{270}, 1639 (1995); R.Held et
  al., Appl. Phys. Lett. \textbf{71}, 2689 (1997).

\bibitem{rokhinson02}
L.~P. Rokhinson, D.~C. Tsui, L.~N. Pfeiffer, and K.~W. West, Superlattices
  Microstruct. {\bf 32},  99  (2002).

\bibitem{heremans94}
J. Heremans, M. Santos, K. Hirakawa, and M. Shayegan, J. Appl. Phys. {\bf 76},
  1980   (1994).

\bibitem{wharam88}
D.~A. Wharam {\it et~al.}, J.~Phys.~C {\bf 21},  L209  (1988).

\bibitem{vanwees88}
B.~J. van Wees {\it et~al.}, Phys. Rev. Lett. {\bf 60},  848  (1988).

\bibitem{zailer94}
I. Zailer {\it et~al.}, Phys. Rev. B {\bf 49},  5101  (1994).

\bibitem{winkler03}
R. Winkler, {\em Spin-Orbit Coupling Effects in Two-Dimensional Electron and
  Hole Systems}, Vol.~191 of {\em Springer Tracts in Modern Physics} (Springer,
  Berlin, 2003).

\bibitem{Salis99}
G. Salis {\it et~al.}, Phys. Rev. B {\bf 60},  7756  (1999).

\end{thebibliography}

\end{document}